# The Ks-band Tully-Fisher Relation - A Determination of the Hubble Parameter from 218 ScI Galaxies and 16 Galaxy Clusters

David G. Russell

Owego Free Academy, Owego, NY 13827 USA

Russeld1@oacsd.org

**Abstract**

The value of the Hubble Parameter ($H_0$) is determined using the morphologically type dependent Ks-band Tully-Fisher Relation (K-TFR).   The slope and zero point are determined using 36 calibrator galaxies with ScI morphology.  Calibration distances are adopted from direct Cepheid distances, and group or companion distances derived with the Surface Brightness Fluctuation Method or Type Ia Supernova.  It is found that a small morphological type effect is present in the K-TFR such that ScI galaxies are more luminous at a given rotational velocity than Sa/Sb galaxies and Sbc/Sc galaxies of later luminosity classes.   Distances are determined to 16 galaxy clusters and 218 ScI galaxies with minimum distances of 40.0 Mpc.   From the 16 galaxy clusters a weighted mean Hubble Parameter of $H_0$=84.2 +/-6 km s$^{-1}$ Mpc$^{-1}$ is found.    From the 218 ScI galaxies a Hubble Parameter of $H_0$=83.4 +/-8 km s$^{-1}$ Mpc$^{-1}$ is found.   When the zero point of the K-TFR is corrected to account for recent results that find a Large Magellanic Cloud distance modulus of 18.39 +/- 0.05 a Hubble Parameter of 88.0 +/- 6 km s$^{-1}$ Mpc$^{-1}$ is found. Effects from Malmquist bias are shown to be negligible in this sample as galaxies are restricted to a minimum rotational velocity of 150 km s$^{-1}$.   It is also shown that the results of this study are negligibly affected by the adopted slope for the K-TFR, inclination binning, and distance binning.   A comparison with the results of the Hubble Key Project (Freedman et al 2001) is made.  Discrepancies between the K-TFR distances and the HKP I-TFR distances are discussed.   Implications for Λ-CDM cosmology are considered with $H_0$=84 km s$^{-1}$ Mpc$^{-1}$.   It is concluded that it is very difficult to reconcile the value of $H_0$ found in this study with ages of the oldest globular clusters and matter density of the universe derived from galaxy clusters in the context of Λ-CDM cosmology.

## Introduction

Hubble (1929) discovered the existence of a linear relationship between the radial velocity (cz) of a galaxy and the distance to the galaxy estimated from absolute magnitude criteria. The Hubble Law is a key component of cosmological theory and the current value of the Hubble Parameter ($H_0$) provides an important constraint on cosmological models (eg. Spergel et al 2003,2006; Hinshaw et al 2009, Dunkley et al 2009; Komatsu et al 2009).

Determination of the Hubble Parameter requires accurate redshift independent distances to a large number of galaxies or clusters of galaxies. The Hubble Key Project (Freedman et al 2001 and references therein – hereafter HKP) utilized several dozen galaxies with Cepheid distances to calibrate five secondary distance indicators – Type Ia SN, Type II SN, the Fundamental Plane (FP), the Surface Brightness Fluctuation Method (SBF), and the Tully-Fisher Relation (TFR). From these five methods the HKP found $H_0$=72 km s$^{-1}$ Mpc$^{-1}$. This value is supported by the recent WMAP results (Spergel et al 2003,2006; Hinshaw et al 2009; Dunkley et al 2009; Komatsu et al 2009).

While there is general agreement that $H_0$=~70 to 74 +/-8 km s$^{-1}$ Mpc$^{-1}$, there have been recent large studies that suggest higher or lower values for the Hubble Parameter. Sandage et al (2006) find $H_0$=62 +/-5 km s$^{-1}$ Mpc$^{-1}$. Tully&Pierce (2000) found $H_0$=77 +/-8 km s$^{-1}$ Mpc$^{-1}$ from 12 galaxy clusters with the I-band TFR. The Tully&Pierce (TP00) study pre-dates the final HKP Cepheid distances (Freedman et al 2001). Utilizing the HKP Cepheid distances with the 24 zero point calibrators in the TP00 study results in a downward revision of their I-band zero point from 21.56 to 21.50 with a resulting increase in $H_0$ to 79 km s$^{-1}$ Mpc$^{-1}$. This value is closer to the value the HKP found with the FP ($H_0$=82 +/-9 km s$^{-1}$ Mpc$^{-1}$) than the final adopted value of $H_0$=72 km s$^{-1}$ Mpc$^{-1}$.

Recent work also argues for a shorter distance to the Large Magellanic Cloud (LMC) than the value adopted by the HKP (Macri et al 2006; van Leeuwen et al 2007; Benedict et al 2007; An et al 2007; Grocholski et al 2007; Catelan&Cortes 2008; Feast et al 2008). These studies suggest a LMC distance modulus of 18.39+/- 0.05 whereas the HKP adopted a LMC distance modulus of 18.50 +/-0.10. The newer LMC distance modulus results in an increase of $H_0$ of ~ 5%.

In this work, the value of $H_0$ is re-evaluated using the Ks-band TFR and taking advantage of recent improvements in data available for Tully-Fisher studies.    The Two Micron All Sky Survey (2MASS – Strutskie et al 2006) has provided near infrared Ks-band photometry for a much larger sample of galaxies than was available for the study of TP00.   The use of the K-band TFR is advantageous because extinction corrections are significantly smaller in the K-band than in the B-band or I-band.    Recently, Springob et al (2007) has corrected for systematic differences between rotational velocities measured from 21cm emission and rotational velocities measured from optical rotation curves (Mathewson&Ford 1996) and provided a large database of uniformly corrected spiral galaxy rotational velocities for Tully-Fisher studies.

Utilizing these data sources and the morphologically type dependent TFR (Russell 2004) strict selection criteria are applied to provide a highly accurate set of distances to 16 galaxy clusters and over 200 ScI galaxies with distances greater than 40 Mpc.   From the galaxy clusters the Hubble parameter is found to be $H_0 = 84.2$ +/-6 km s$^{-1}$ Mpc$^{-1}$. The ScI galaxy sample gives $H_0 = 83.4$ +/-8 km s$^{-1}$ Mpc$^{-1}$.

This paper is organized as follows:   Section 2 describes the calibration of the morphologically type dependent K-band TFR.   Section 3 describes the sample selection for determination of the Hubble Parameter.   Section 4 discusses the value of the Hubble Parameter found in this study and considers possible influences on the value of $H_0$. Section 5 compares the results of this study with those of the HKP.   Section 6 is a discussion of implications for cosmology if $H_0 = 84$ km s$^{-1}$ Mpc$^{-1}$ and a brief conclusion is provided in section 7.

**2. Calibration, sample selection, and scatter of the K-band TFR for ScI galaxies**

*2.1 Calibration of $K_s$-Band TFR*

The accuracy of results derived from TFR distance estimates depends critically upon a large calibration sample for determination of the slope and zero point and strict selection criteria that eliminate galaxies from the sample which are most likely to have large TFR errors.    The following sections describe the selection criteria utilized in the calibration of the $K_s$-TFR for this analysis.

### 2.1.1  Morphological Type effect in the TFR

Russell (2004, 2005a) found evidence that the B-band TFR may be split into two morphological groups with identical slope but a zero-point offset of 0.57 mag.   The first morphological group (ScI group) includes galaxies classified as ScI, ScI-II, ScII, SbcI, SbcI-II, SbcII and Seyfert galaxies with spiral morphology.   The second morphological group (Sb/ScIII group) includes Sab/Sb/Sbc/Sc galaxies that do not have ScI group morphology.   ScI group galaxies are more luminous at a given rotational velocity than Sb/ScIII group galaxies (Russell 2004).   Failure to account for this type effect results in overestimated B-band TFR distances for Sb/ScIII group galaxies, underestimated B-band TFR distances for ScI group galaxies, and a TFR slope that is too steep.   While the morphological type effect is largest in the B-band, it persists in the I-band and Ks-band (Giovanelli et al 1997b; Masters et al 2008)

For calibration of the $K_s$-band TFR (hereafter K-TFR) in this analysis the ScI group calibrators were restricted to galaxies with Hubble T-types ranging from 3.1 to 6.0 and with luminosity classes ranging from 1.0 to 3.9 as classified in HyperLeda (Paturel et al 2003).   The ScI group was calibrated, and then cluster samples were used to determine the zero point offset for the Sb/ScIII group as discussed in section 2.2.   The ScI group calibrator sample includes 10 galaxies with direct distance determinations from Cepheid variables (Freedman et al 2001); 1 galaxy (NGC 2903) with a direct distance determined from photometry of bright stars (Drozdovsky&Karachentsev 2000); 18 galaxies that are members of groups or companions of galaxies from the SBF survey of Tonry et al (2001); and 7 galaxies that are companions of galaxies with Type Ia SN distances from Freedman et al 2001.   As found by Ajhar et al (2001), the SBF distance moduli from Tonry et al (2001) are reduced by 0.06 mag in order to align them with the final HKP Cepheid distance scale.   The calibrator sample is listed in Tables 1 and 2.

### 2.1.2  Rotational velocities

The greatest source of uncertainty in TFR distances is found in the measurement and correction of rotational velocities derived from HI linewidths and optical rotation curves (Haynes et al 1999).   In addition to uncertainty in raw 21cm linewidths, corrections must be made for inclination and turbulence. Inclination corrections are larger and subject to greater uncertainty as galaxies approach face on orientation. For this reason, calibrator galaxies selected for this analysis were restricted to inclinations >30°and ≤ 80°. An inclination limit of 80° was chosen because luminosity class cannot be assigned for edge on galaxies and therefore edge on galaxies cannot be accurately classified as ScI or Sb/ScIII group.

The influence of turbulence corrections is greater at smaller linewidths (Giovanlli et al 1997b). It is also now well established that slower rotating galaxies have larger TFR scatter than faster rotators (Federspiel et al 1994; Giovanelli 1996; Giovanelli et al 1997b; Masters et al 2006, 2008). For example Federspiel et al found that TFR scatter decreases from +/-0.90 mag for the slowest rotators to only +/-0.43 mag for the fastest rotators. Giovanelli (1996) also found that the fastest rotators had a scatter about a factor of 2 smaller than the slowest rotators.

In order to avoid the problems associated with the large TFR scatter of the slowest rotators, calibrators were also restricted to galaxies with rotational velocities ≥ 150 km s$^{-1}$ in the Springob et al (2007) rotational velocity database.

### 2.1.3 Ks-band magnitudes and corrections

2MASS total $K_s$-band magnitudes (Strutskie et al 2006) were extracted from NED and corrected for galactic and internal extinction following Tully et al (1998) and a small cosmological k-correction was added following Poggianti (1997). The galactic and internal extinction corrections were made for each galaxy using the B-band extinction corrections provided in Hyperleda (Paturel et al 2003). The corrections are as follows: the galactic B-band extinction (ag in HyperLeda) was multiplied by 0.086 to derive the extinction in the $K_s$-band and the internal absorption correction (ai in HyperLeda) was multiplied by 0.15 to derive the extinction in the $K_s$-band. Finally the k-correction was approximated as -1.52z as per Poggianti (1997). Thus the total corrected $K_s$-band magnitudes (Ktc) in this study were derived from uncorrected 2MASS total $K_s$-band magnitudes according to the following:

$$Ktc = Ktot - 0.086ag - 0.15ai - 1.52z \qquad (1)$$

### 2.1.4 Slope and zero point of the ScI group $K_s$-band TFR

Figure 1 is a plot of the absolute $K_s$-band magnitude ($M_K$) versus the logarithm of the rotational velocity for the 36 ScI group calibrators listed in Tables 1 and 2. All rotational velocities used in this study are drawn from the sample of Springob et al (2007). The solid line in Figure 1 is a least squares fit and has a slope of -8.22 +/- 0.37. The value of the slope does not significantly change when determined independently for the various subsamples from which the calibration distances were drawn. The 11 Table 1 calibrators with direct distance estimates, 18 Table 2 calibrators with SBF distance estimates, and 7 Table 2 calibrators with Type Ia SN distance estimates give slopes of -7.78, -8.51, and -8.16 respectively.

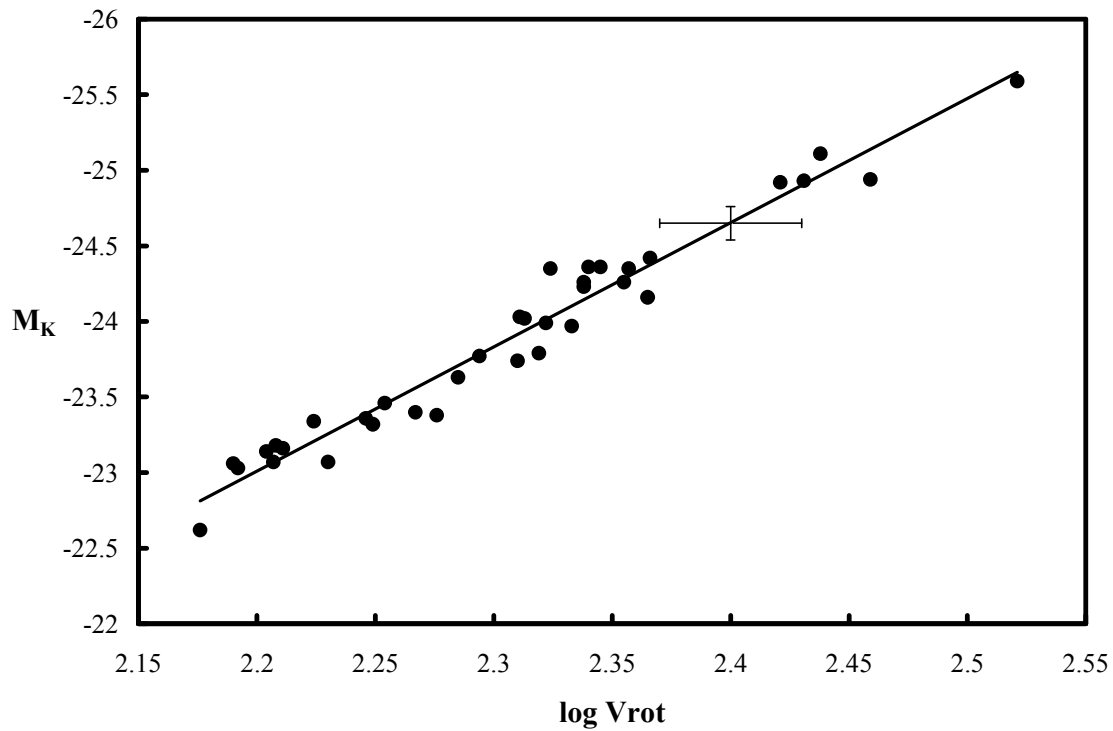

*Figure 1 – Calibration of the K-TFR with 36 calibrators from Tables 1 and 2. Solid line is a slope of -8.22 and error bars represent typical uncertainty in log Vrot and $M_K$.*

Table 1:  Calibrators with direct distance measurements

| Galaxy | log Vrot | +/- | *incl* ° | Ktc | m-M | zp $K_s$ |
|--------|----------|-----|----------|-----|-----|----------|
| N1365 | 2.459 | .088 | 39 | 6.33 | 31.27 | 22.81 |
| N1425 | 2.254 | .021 | 62 | 8.24 | 31.70 | 23.02 |
| N2903 | 2.319 | .012 | 60 | 5.96 | 29.75 | 22.81 |
| N3198 | 2.190 | .009 | 71 | 7.64 | 30.70 | 23.14 |
| N3351 | 2.267 | .036 | 44 | 6.61 | 30.01 | 22.85 |
| N4321 | 2.357 | .079 | 32 | 6.56 | 30.91 | 23.06 |
| N4414 | 2.340 | .038 | 51 | 6.88 | 31.24 | 23.21 |
| N4535 | 2.285 | .037 | 45 | 7.36 | 30.99 | 22.93 |
| N4548 | 2.333 | .040 | 35 | 7.08 | 31.05 | 22.87 |
| N4725 | 2.324 | .030 | 63 | 6.11 | 30.46 | 23.33 |
| N7331 | 2.431 | .006 | 65 | 5.91 | 30.84 | 23.03 |

Using the slope of -8.22 derived from the full sample of calibrators, a mean zero point of 23.01+/-0.14 is found for the 36 calibrator galaxies. With the slope of -8.22, the value of the zero point is not affected by the method from which the calibration distance was determined.  The 11 galaxies with direct distance determinations in Table 1, 18 galaxies with SBF distance determinations in Table 2, and 7 galaxies with Type Ia SN distance determinations in Table 2 give mean zero points of 23.01 +/-0.17, 23.01+/-0.14, and 23.02 +/-0.06 respectively.

K-TFR distance moduli for ScI group galaxies in this analysis are calculated using the following equation:

$$m-M_K = Ktc + 8.22(logVrot - 2.2) + 23.01 \qquad (2)$$

It is important to note that equation 2 should not be viewed as the global K-band TFR. The sample has a lower rotational velocity limit of 150 km s$^{-1}$, morphological restriction to Hubble T-types 3.1 to 6.0, and luminosity class restriction to luminosity classes I, I-II, II, and II-III.  Caution should be used in applying the calibration developed here to galaxies that fall outside these ranges.

Table 2:   Calibrators with companion SBF or SN Ia distance measurements

| Galaxy | log Vrot | +/- | *incl* ° | Ktc | m-M | zp $K_s$ | Companion |
|--------|----------|-----|----------|-----|-----|----------|-----------|
| N3089 | 2.246 | 0.019 | 52 | 9.24 | 32.60 | 22.98 | Antlia |
| N3095 | 2.313 | 0.043 | 55 | 8.58 | 32.60 | 23.09 | Antlia |
| N3223 | 2.438 | 0.021 | 47 | 7.49 | 32.60 | 23.15 | Antlia |
| N3318 | 2.310 | 0.034 | 58 | 8.86 | 32.60 | 22.84 | Antlia |
| N3347 | 2.338 | 0.048 | 69 | 8.34 | 32.60 | 23.13 | Antlia |
| I2560 | 2.311 | 0.016 | 64 | 8.57 | 32.60 | 23.12 | Antlia |
| I2522 | 2.276 | 0.043 | 48 | 9.22 | 32.60 | 22.76 | Antlia |
| N4575 | 2.224 | 0.027 | 51 | 9.24 | 32.58 | 23.14 | Cen30 |
| N4603 | 2.345 | 0.007 | 51 | 8.25 | 32.58 | 23.14 | Cen30 |
| N1255 | 2.204 | 0.038 | 45 | 8.33 | 31.47 | 23.11 | N1201 |
| N3054 | 2.366 | 0.017 | 54 | 8.25 | 32.67 | 23.06 | N3078 |
| N5011a | 2.192 | 0.037 | 57 | 10.02 | 33.05 | 23.10 | N5011 |
| N5033 | 2.365 | 0.019 | 63 | 6.87 | 31.03 | 22.80 | N5273 |
| N7610 | 2.176 | 0.039 | 52 | 10.94 | 33.56 | 22.82 | N7619 |
| I4538 | 2.208 | 0.016 | 39 | 9.41 | 32.59 | 23.11 | N5903 |
| E582-12 | 2.211 | 0.020 | 56 | 9.10 | 32.26 | 23.07 | N5898 |
| E377-31 | 2.230 | 0.015 | 60 | 10.17 | 33.24 | 22.82 | N3557 |
| E287-13 | 2.249 | 0.017 | 78 | 9.17 | 32.49 | 22.92 | N7097 |
| E471-49 | 2.521 | 0.013 | 60 | 9.80 | 35.39 | 22.95 | E471-27 |
| E471-51 | 2.322 | 0.036 | 53 | 11.40 | 35.39 | 22.99 | E471-27 |
| E471-2 | 2.355 | 0.012 | 71 | 11.13 | 35.39 | 22.99 | E471-27 |
| E577-1 | 2.294 | 0.024 | 71 | 11.28 | 35.05 | 23.00 | E508-67 |
| N4541 | 2.421 | 0.019 | 64 | 10.01 | 34.93 | 23.10 | N4493 |
| E444-31 | 2.338 | 0.020 | 50 | 11.11 | 35.34 | 23.10 | I4232 |
| A530465 | 2.207 | 0.021 | 67 | 12.27 | 35.34 | 23.01 | I4232 |

*2.2  Morphological Type dependence in the K-band TFR*

Russell (2004) found that ScI group galaxies have a zero point 0.57 mag larger than Sb/ScIII galaxies.    The effect of morphological type dependence is known to be smaller in the I-band where Giovanelli et al (1997a) found a 0.32 mag smaller zero point for Sa/Sab galaxies and 0.10 mag smaller zero point for Sb galaxies relative to Sbc and later type spirals.    It has generally been thought that any type effects should disappear in the near infrared bands.    However, Masters et al (2008) recently found evidence that there is a small type effect in the J,H, and K-band Tully-Fisher relations.

The Masters et al (2008) sample includes a much broader range of morphological types (eg. extreme late type and early type spirals) and slower rotators that have been excluded from this study. In order to test for a morphological type effect in the K-band we define the Sb group galaxies as all non-Seyfert galaxies of morphological T-types 1.0 to 3.0 and as found by Russell (2004) include in the Sb group later type spirals of Hubble T-types 3.1-6.0 and luminosity classes $\geq 4.0$ as classified in HyperLeda.

If a type effect similar to that found in the B-band TFR (Russell 2004) exists in the K-TFR, then Sb galaxies within clusters should have larger mean distances than ScI group galaxies when Sb group distances are calculated using the ScI group zero point. To test for this effect, clusters were selected from the template sample of Springob et al (2007). The selection criteria adopted for the galaxies within the clusters was consistent with that utilized for the ScI group calibrator sample. Specifically, galaxies were required to have inclinations between 35° and 80°, rotational velocities $\geq 150$ km s$^{-1}$, and Hubble T-types from 1.0 to 6.0. Clusters used for estimating the size of the type effect in the K-band all had at least 12 member galaxies from the Springob et al sample meeting the above criteria and a minimum of 4 galaxies in both the ScI and the Sb morphological groups. For the Abell 400 cluster the sample was also restricted to galaxies with redshifts ranging from 6500 km s$^{-1}$ to 8000 km s$^{-1}$.

At larger distances, fewer galaxies in the HyperLeda database are assigned luminosity classes. For this reason, before calculating distances the HyperLeda image of each candidate galaxy meeting all other criteria was visually inspected in order to determine if the morphology of the galaxy is ScI group morphology or Sb group morphology. For example NGC 551(Figure 2a) in the Pisces supercluster is classified as SBbc in HyperLeda, but no luminosity class is provided. Visual inspection of the DSS image in HyperLeda confirms NGC 551 has arm structure of ScI group morphology and thus is included in the ScI group sample.

Visual inspection also serves as an independent check on the inclination estimates provided by Springob et al (2007). For example, Springob et al report an inclination of 51° for NGC 2582 in the Cancer cluster. However, visual inspection of the NGC 2582 image (Figure 2b) clearly indicates an inclination much closer to face on orientation than 51°. UGC 1695 in the Pisces supercluster has an inclination of 76° in the Springob et al database and thus would meet the selection criteria of this study.    However, visual inspection of the UGC 1695 image (Figure 2c) reveals an inclination very close to 90° and it is therefore impossible to determine the luminosity class of this galaxy.  Galaxies for which visual inspection of the image indicated an unambiguous problem with the reported inclination were rejected from the sample.

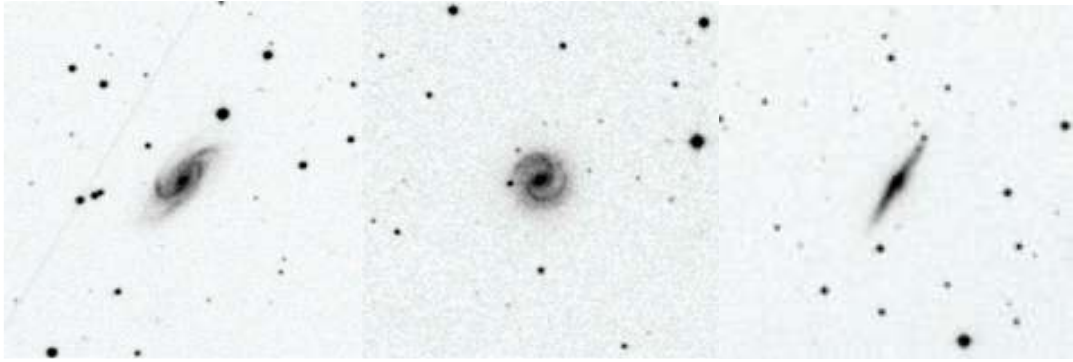

*Figure 2a*          *Figure 2b*          *Figure 2c*

The Pisces Supercluster (Abell 262, NGC 507, and NGC 383 clusters), Abell 400, Coma, Cancer, Abell 1367, and Hydra clusters had enough galaxies meeting the selection criteria above to estimate the size of the morphological type effect in the K-band. Distances to all galaxies not removed from the sample by the selection criteria described above were calculated for these 6 clusters using equation 2.  The resulting mean distances to the Sb group and ScI group galaxies in each cluster are provided in Table 3.    It can be seen that for each cluster the mean Sb group distance to the cluster is greater than the mean ScI group distance to the cluster by +0.06 to +0.40 mag.   The mean difference for the 56 Sb group galaxies in Table 3 is +0.19 +/-0.10 mag.   This indicates that a smaller but still detectable type effect remains in the K-band TFR.  All distances to Sb group galaxies will therefore be calculated using equation 2, but with a zero point of 22.82 rather than the zero point of 23.01 found for ScI group galaxies.

Table 3:   Type Effect in the K-band TFR

| Cluster | N Sb | m-M Sb | N ScI | m-M ScI | Sb - ScI |
|---------|------|--------|-------|---------|----------|
| Pisces  | 13   | 34.02  | 17    | 33.87   | +0.15    |
| A400    | 10   | 34.84  | 6     | 34.58   | +0.26    |
| Coma    | 9    | 34.65  | 6     | 34.59   | +0.06    |
| Cancer  | 9    | 34.21  | 6     | 33.81   | +0.40    |
| A1367   | 11   | 34.72  | 4     | 34.60   | +0.12    |
| Hydra   | 4    | 33.84  | 8     | 33.49   | +0.35    |

*2.3 Comparison with other K-band TFR studies*

One of the largest potential sources of a systematic bias in TFR distances arises when an incorrect TFR slope is applied to the TFR sample.  For this reason it is important that the  TFR derived in this study is only applied to galaxies meeting the selection criteria utilized to calibrate the type dependent K-band TFR.

Karachentsev et al (2002), Noordermeer and Verheijen (2007), and Masters et al (2008) developed K-band Tully-Fisher relations from 2MASS photometry that are suitable for comparison with this study.    Karachentsev et al found a K-TFR slope of -9.02 +/-0.25 using Hubble distances for 436 galaxies.    No corrections for morphological type dependence are applied and the sample includes galaxies with rotational velocities as small as 75 km s$^{-1}$.   Noordmeer & Verheijen found a K-TFR slope of -8.65 +/- 0.19 from a sample of 48 spirals with rotational velocities as small as 83 km s$^{-1}$.  No corrections for morphological type dependence were applied.   Masters et al used a "basket of clusters" technique to create a K-TFR template with 888 spiral galaxies including galaxies with rotational velocities smaller than 75 km s$^{-1}$.    The Masters et al sample provides a global K-TFR corrected for several types of bias (see section 3 of Masters et al) with a direct slope of -8.85 for the full sample.

It is interesting to note that these studies utilize different samples and calibration techniques and yet find consistent slopes in the small range of -8.65 to -9.02 for the direct K-band TFR.   These slopes are slightly steeper than the slope of -8.22 found in this study for a sample restricted to galaxies with ScI group morphology and with rotational velocities $\geq$ 150 km s$^{-1}$.    Masters et al (2008) also applied all bias corrections and the direct TFR to 374 "High mass" galaxies in their template sample with rotational velocities in excess of 160 km s$^{-1}$ and found a shallower slope of  -8.06 which is consistent with the slope found with the 36 calibrators in Tables 1 and 2.   This suggests that the shallower slope found in this study results from the exclusion of slower rotators from the sample.

The slope for the ScI calibrator sample was also determined using only the 33 ScI galaxies with rotational velocities $\geq$ 160 km s$^{-1}$ and was found to remain unchanged.  It is reassuring to note that the slope derived from 36 ScI group calibrators is only slightly steeper than the slope Masters et al found from 374 comparable galaxies after applying their complete set of bias corrections. The slope found in this study is unbiased as long as it is applied to galaxies meeting the selection criteria utilized in compiling the calibrator sample. Inclusion of slower rotating galaxies would require a steeper slope.

Masters et al (2008) also calibrated the K-TFR using a bivariate fit and found an even steeper slope of -10.02 for their complete sample of 888 galaxies. In order to test for the possible influence of the slope on the results of this study the zero point was calculated for the 36 ScI group calibrators using a slope of -10.02. The resulting zero point is reduced from 23.01 +/-0.14 to 22.84 +/- 0.20 with the steeper slope. Keeping a 0.19 mag type effect, Sb group galaxy distances will be calculated using a zero point of 22.65 when the slope of -10.02 is utilized.

In this analysis, the Hubble Parameter will be derived from the slope and zero points found in sections 2.1.4 and 2.2. However, as a test of the effect of the K-TFR slope on the Hubble Parameter, the distances to galaxies in the cluster sample (section 3.1) will also be calculated using the slope of -10.02 and the zero points of 22.84 and 22.65 for the ScI and Sb group respectively(section 4.2).

*2.4 Scatter in the Type dependent K-band TFR*

The subject of scatter in the TFR has been widely studied (Bernstein et al 1994,Willick 1996, Raychaudhury et al 1997, Giovanelli et al 1997b, Tully&Pierce 2000, Sakai et al 2000, Kannappan et al 2002, Russell 2005b, Masters et al 2008). For the purposes of this study it is important to note that the scatter observed with this sample is only applicable to the selection criteria utilized in creating the Tully Fisher relations discussed in sections 2.1 and 2.2. Larger intrinsic scatter will occur with a less restrictive set of sample selection criteria. The RMS scatter for the 36 ScI group calibrators in Tables 1 and 2 is +/- 0.14 mag relative to the calibration distance moduli.

Most 2MASS $K_s$-band magnitudes for galaxies in the sample have an uncertainty of only +/-0.03 to +/-0.10 mag. However, Noordermeer&Verheijen (2007) point to reasons for suspecting that the 2MASS uncertainty estimates are overly optimistic and suggest a more realistic estimate of +/-0.11 mag for the typical K-band uncertainty – which is adopted for this study. The mean uncertainty of logarithm of the rotational velocities (Springob et al 2007) used for the 36 calibrators is +/- 0.030. Thus the greatest source of uncertainty in the K-band TFR distances should be expected to arise from the effect of inclination uncertainty on the correction of the rotational velocity to edge on orientation. Since visual inspection of images eliminated galaxies with grossly inaccurate or uncertain inclinations (section 2.2), the galaxies in the final sample have inclinations that should be accurate to 5° or better.

With a slope of -8.22 the typical distance modulus uncertainty is +/-0.247 mag. from the uncertainty in the rotational velocities as reported by Springob et al (2007). This may be added in quadrature with a $K_s$-band magnitude uncertainty of +/-0.11 mag to yield a total expected distance modulus uncertainty of +/-0.27 mag. The RMS scatter of the ScI group calibrator zero point is only +/- 0.14 mag which is significantly smaller than the scatter expected from the rotational velocities. This may be an indication that the uncertainty in the rotational velocities is overestimated by Springob et al. The mean uncertainty Springob et al reported for the rotational velocities is of the magnitude expected for a +/- 6° inclination uncertainty. The small observed scatter suggests that inclinations are actually accurate to ~+/- 3° in most cases. In any case, the small observed scatter in the K-TFR zero point indicates that there is negligible intrinsic scatter in the K-TFR for galaxies meeting the selection criteria adopted in this study.

## 3. Samples for determination of the Hubble Parameter

### 3.1 Springob et al galaxy cluster sample

Table 4 lists the 10 clusters from the template sample of Springob et al (2007) that had at least 5 galaxies meeting the selection criteria utilized to define the calibration samples as discussed in sections 2.1 and 2.2. Column 1 is the cluster. Column 2 is the number of cluster members in the full Springob et al sample. Columns 3,4, and 5 are the number of galaxies within each cluster rejected for not meeting the rotational velocity, inclination, and Hubble T-type criteria respectively. Column 6 is the number of galaxies rejected for having K-TFR distances that fall outside the primary distribution of K-TFR distances for the cluster. Column 7 provides notes on several rejections specific to individual clusters or galaxies. Column 8 is the final sample of accepted galaxies for each cluster. In order to reduce the effect of motions within the local velocity field and cluster depth effects, the cluster sample was also restricted to those clusters with a minimum distance of 40.0 Mpc.

In addition to the criteria discussed in section 2, galaxies were restricted to the redshift range of 6500-8000 km s$^{-1}$ for the Abell 400 cluster and 8000-10000 km s$^{-1}$ for the Abell 2197/99 cluster. Six of the 10 clusters had member galaxies meeting all other criteria rejected based upon the distance distribution of the cluster members (Column 6 of Table 4). With the exception of UGC 8229 in the Coma cluster which has a K-TFR distance modulus 2.3σ less than the Coma cluster mean, all galaxies rejected as being outside the primary cluster distance distribution were at least +/-2.5σ from the mean of the accepted galaxies.

The largest sample of galaxies from the Springob et al template sample is found in the Pisces supercluster which is comprised of galaxies in the Abell 262, NGC 507, and NGC 383 clusters. The selection criteria utilized in this study is illustrated using this sample of 95 galaxies (Table 4). As indicated in Table 4, 23 galaxies from the Pisces supercluster met all selection criteria applied. These 23 galaxies have a mean distance modulus of 33.77 +/-0.21 and all have distance moduli within the range 33.35 to 34.19.

Table 4: Cluster template sample rejection reasons

| Cluster | N | log Vrot | i | t | Distance Distribution | other | Final Sample |
|---------|---|----------|---|---|-----------------------|-------|--------------|
| Pisces | 95 | 39 | 19 | 7 | 7 | | 23 |
| A400 | 50 | 13 | 7 | 2 | 4 | 3 declination rejected 9 redshift range rejected | 12 |
| Cancer | 49 | 20 | 13 | 1 | 5 | | 10 |
| Coma | 43 | 14 | 9 | 5 | 4 | | 11 |
| A1367 | 33 | 6 | 6 | 6 | 3 | | 12 |
| Hydra | 31 | 9 | 9 | 0 | 2 | 1 uncertain 2MASS k$_s$ | 10 |
| A3574 | 29 | 12 | 8 | 0 | 0 | 1 no ks; 1 R.A. rejected | 7 |
| A2197/99 | 22 | 1 | 8 | 0 | 0 | 4 redshift range rejected | 9 |
| A2634 | 22 | 9 | 3 | 2 | 0 | | 8 |
| A779 | 17 | 7 | 2 | 4 | 0 | | 5 |

Note that 39 of the 95 galaxies in the Pisces supercluster template sample had rotational velocities less than 150 km s$^{-1}$. For 27 of these galaxies it was possible to calculate K-TFR distances. The 27 slow rotators had a mean distance modulus of 34.06 with a substantially larger RMS scatter of +/-0.87 mag when compared with the 23 accepted galaxies. In fact only 8 of the 27 slow rotators have K-TFR distance moduli that fall within the distance modulus range of the 23 accepted galaxies. This situation does not improve if the slope of -10.02 is adopted as the mean distance for the slow rotators is then decreased to 33.57+/-0.90. The significantly larger scatter found for the slow rotators justifies the exclusion of slower rotators from the sample and is consistent with the findings of Federspiel et al (1994) who found a scatter of +/- 0.90 mag for slow rotators but only +/- 0.40 mag for faster rotators. It was also suggested by Giovanelli (1996) that the safest means of obtaining the tightest possible TFR is to exclude lower luminosity galaxies.

The Pisces supercluster sample had 7 galaxies (listed in Table 5) rejected for a distance modulus that deviated from the mean of the accepted galaxies by +/- 2.5σ or greater. Accepting these 7 galaxies would slightly increase the mean distance modulus to 33.86 but significantly increase the observed RMS scatter about the mean to +/-0.45 mag. The accepted galaxies have K-TFR distances ranging from 47 to 69 Mpc. It seems prudent then to reject from the Pisces supercluster sample a galaxy such as UGC 1416 which has a K-TFR distance of 104 Mpc and therefore potentially a large error in the K-TFR distance. In fact UGC 1416 may be a genuine background galaxy as the 2MASS total K-band angular diameter of UGC 1416 (rotational velocity = 209 km s$^{-1}$) is only 1.00′ whereas the accepted Pisces supercluster galaxy UGC 1676 (Vrot = 214 km s$^{-1}$) has a K-band angular diameter of 2.19′. However, whether UGC 1416 is a genuine background galaxy or a galaxy with an extremely large K-TFR distance error, it would be inappropriate to use UGC 1416 in calculating the mean distance to the Pisces supercluster as it is clearly not representative of the normal distance distribution of cluster members.

The 10 Springob et al template clusters were supplemented with 6 additional clusters or groups from the non-template sample of Springob et al which had at least 5 galaxies meeting the selection criteria of this study. The value of the Hubble Parameter derived from these 16 clusters is discussed in section 4.

Table 5:  Pisces supercluster galaxies rejected for distance distribution

| Galaxy | m-M K-TFR | Vcmb | Δm-M | Mpc |
|--------|-----------|------|------|-----|
| NGC 688 | 34.36 | 3894 | +0.59 | 74.5 |
| UGC 1744 | 34.67 | 4578 | +0.90 | 85.9 |
| U1416 | 35.08 | 5222 | +1.31 | 103.8 |
| U1257 | 34.59 | 4404 | +0.82 | 82.8 |
| U724 | 34.63 | 4880 | +0.86 | 84.3 |
| U1493 | 33.05 | 3895 | -0.72 | 40.8 |
| U1672 | 32.96 | 4367 | -0.81 | 39.1 |

*3.2   ScI galaxies from the Mathewson&Ford Sample*

Mathewson & Ford (1996 – MF96) provided a catalog of 2447 galaxies- most with rotational velocities derived from optical rotation curves.   Springob et al (2007) corrected the rotational velocities from MF96 to standardize them with rotational velocities derived from hydrogen linewidths and have included the MF96 sample in their non-template catalog.   The MF96 catalog was searched for all ScI group galaxies meeting the rotational velocity, inclination, and morphology selection criteria of this study.   This yielded 140 galaxies with ScI group morphology as classified in HyperLeda and distances of at least 40.0 Mpc but no greater than 140.0 Mpc.

The image of each galaxy from the MF96 sample with Hubble T-types from 3.1 -6.0 but for which the luminosity class was not provided was visually inspected.   An additional 78 ScI group galaxies were identified by examining arm structure in the images bringing the total sample to 218 ScI group galaxies with distances in the 40.0 Mpc to 140.0 Mpc range.   Note that these 78 galaxies were identified visually before determining the K-TFR distance and therefore the decision of whether or not to include these galaxies in the sample was not influenced by prior knowledge of the distance to each galaxy.

In the process of identifying the 218 ScI group galaxies, 12 galaxies were identified for which there was a 15 degree or greater discrepancy between the HyperLeda and the MF96 inclination, or for which the inclination provided in MF96 was highly uncertain. In some cases the inclination uncertainty results from unusual elongated arm structure that produces an inclination closer to edge on orientation than visual inspection of the image indicates.   For example, ESO 384-9 has a 59 degree inclination in Springob et al but the visual appearance suggests an inclination closer to 40 degrees with arms that are significantly elongated. Visual inspection of the 218 accepted ScI group galaxies confirmed that the Springob et al inclinations for the accepted galaxies are reasonable.

For these 218 ScI galaxies, rotational velocities were taken from Springob et al and 2MASS Ks-band magnitudes were corrected as discussed in section 2.1.3. The value of the Hubble Parameter derived from the 218 ScI galaxies is discussed in the next section.

## 4. The Hubble Parameter

### 4.1 The value of the Hubble Parameter from 16 galaxy clusters and 218 ScI galaxies

Table 6 lists the mean distances, redshifts and the value of the Hubble parameter indicated for each of the 16 galaxy clusters discussed in section 3.1. The mean RMS scatter of the distance moduli of cluster members around the cluster mean is +/-0.26 mag. Note that the intrinsic K-TFR scatter must be smaller than this because of an expected contribution to the observed scatter from cluster depth effects. The 10 template clusters give an unweighted mean for the Hubble Parameter of 85.1 +/-5 km s$^{-1}$ Mpc$^{-1}$. The weighted mean is 85.0+/-5 km s$^{-1}$ Mpc$^{-1}$. The individual template clusters give values of the Hubble Parameter within the remarkably small range from $H_0$=82.9 km s$^{-1}$ Mpc$^{-1}$ to $H_0$=86.9 km s$^{-1}$ Mpc$^{-1}$. The 6 non-template clusters generally have smaller numbers of accepted cluster members and have a larger range in $H_0$ values ($H_0$=70.1 to 93.3). The unweighted mean for the full sample of 16 clusters is $H_0$= 84.9+/-5 km s$^{-1}$ Mpc$^{-1}$ and the weighted mean is $H_0$=84.2+/-5 km s$^{-1}$ Mpc$^{-1}$. The redshift velocity – distance relation for the 16 clusters is shown in Figure 3.

Table 7 lists the mean values of $H_0$ grouped into 5 distance bins for the 218 ScI galaxies discussed in section 3.2. The mean value of $H_0$ for the 218 ScI galaxies is 83.4 +/-8 km s$^{-1}$ Mpc$^{-1}$ which is very close to the weighted mean value found for the 16 clusters in Table 6. The two samples therefore indicate a Hubble Parameter of 84 +/-6 km s$^{-1}$ Mpc$^{-1}$ using the morphologically type dependent K-TFR.

It is important to note that if Malmquist bias significantly affects the sample, then the value of the Hubble Parameter is expected to increase significantly with distance

Table 6:   Cluster Hubble Parameter

| Cluster | N | m-M | +/- | Mpc | Vcmb | $H_0$ | +/- |
|---------|---|-----|-----|-----|------|-------|-----|
| Template | | | | | | | |
| Pisces | 23 | 33.77 | 0.21 | 56.8 | 4794 | 84.4 | 7.7 |
| A1367 | 12 | 34.49 | 0.22 | 79.1 | 6559 | 82.9 | 7.9 |
| A400 | 12 | 34.60 | 0.18 | 83.2 | 7227 | 86.9 | 7.0 |
| Coma | 11 | 34.70 | 0.26 | 87.1 | 7563 | 86.8 | 9.8 |
| Cancer | 10 | 33.61 | 0.34 | 52.7 | 4523 | 85.8 | 12.5 |
| Hydra | 10 | 33.43 | 0.19 | 48.4 | 4080 | 84.3 | 7.0 |
| A2197/99 | 9 | 35.26 | 0.21 | 112.7 | 9479 | 84.1 | 7.8 |
| A2634 | 8 | 35.00 | 0.22 | 100.0 | 8689 | 86.9 | 8.4 |
| A3574 | 7 | 33.80 | 0.29 | 57.6 | 4903 | 85.1 | 10.6 |
| A779 | 5 | 34.76 | 0.43 | 89.5 | 7532 | 84.2 | 15.2 |
| Non-Template | | | | | | | |
| A548 | 8 | 35.64 | 0.43 | 134.3 | 12062 | 89.8 | 16.1 |
| A114 | 7 | 36.52 | 0.26 | 201.4 | 17810 | 88.4 | 9.9 |
| A1736 | 7 | 35.57 | 0.22 | 130.0 | 11585 | 89.1 | 8.6 |
| A3716 | 7 | 35.94 | 0.28 | 154.2 | 14382 | 93.3 | 11.3 |
| ESO 596 | 6 | 34.85 | 0.23 | 93.3 | 7043 | 75.5 | 7.6 |
| ESO 471 | 5 | 35.40 | 0.21 | 120.2 | 8429 | 70.1 | 6.4 |

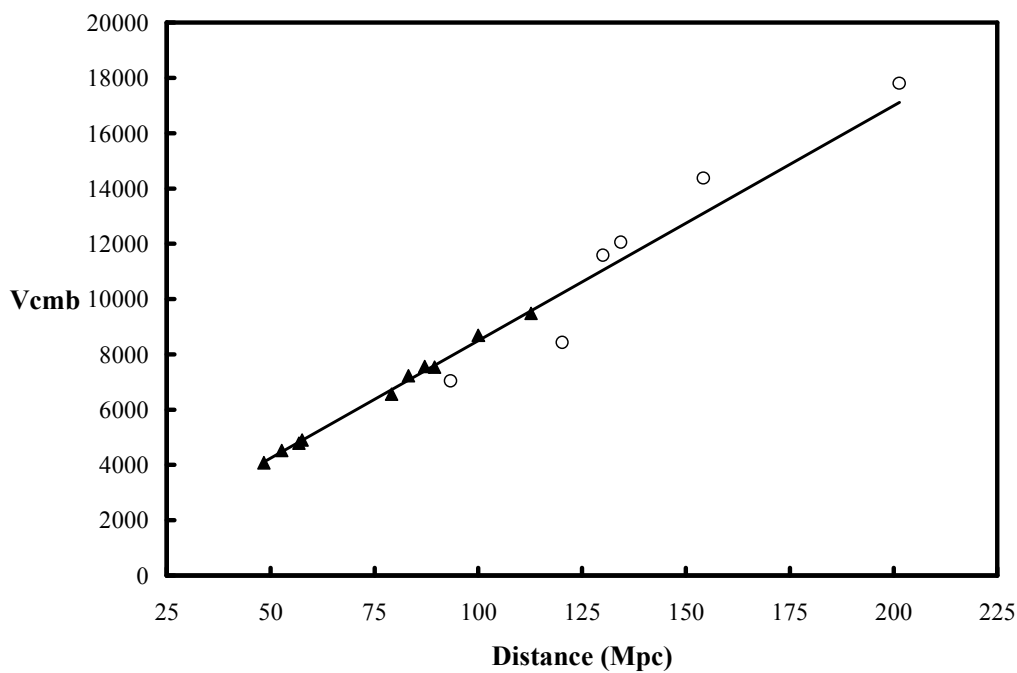

*Figure 3 – Hubble plot for 16 galaxy clusters in Table 6. Filled triangles are template clusters and open circles are non-template clusters.  Solid line represents $H_0$=84.*

(Bottinelli et al 1986,1988; Federspiel et al 1994).  The effects of Malmquist bias are not observed in Table 7 as there is very little variation in $H_0$ from one distance bin to the next.   It is also worth noting that the two most distant distance bins have the lowest mean $H_0$ values and thus run counter to what is expected when there is a significant Malmquist bias.    The lack of a signature from Malmquist bias is not surprising given that the observed scatter of the K-TFR in this study is very small and that Masters et al (2008) concluded that the effects of Malmquist bias are negligible in a sample that included many much slower rotators theoretically more susceptible to introducing Malmquist bias effects than the fast rotators used in this study.

Table 7:   MF96 ScI Hubble parameter

| Distance range | N | log Vrot | $H_0$ |
|---|---|---|---|
| 40-59.9 | 52 | 2.306 | 82.9 |
| 60.0-79.9 | 63 | 2.298 | 83.3 |
| 80.0-99.9 | 52 | 2.326 | 86.5 |
| 100.0-119.9 | 28 | 2.336 | 82.1 |
| 120.0-139.9 | 23 | 2.368 | 79.5 |
| Total | 218 | | 83.4 |

*4.2   Effect of the slope of the K-TFR*

It is worth investigating whether or not the results would change if a steeper slope is adopted for the K-TFR.   As discussed in section 2.3 while Masters et al (2008) find a slope for fast rotators consistent with that found from the 36 calibrators in Tables 1 and 2, they found a slope of -10.02 for the K-TFR from a bivariate fit that includes slower rotators.    Using this global K-TFR slope of -10.02 and the mean zero points for ScI group and Sb group galaxies calculated in section 2.3, distances to all 16 clusters were recalculated (Table 8).    The unweighted mean values for the Hubble parameter are 83.9 +/-7 km s$^{-1}$ Mpc$^{-1}$ for the 10 template clusters and 83.7 +/-7 km s$^{-1}$ Mpc$^{-1}$ for the full sample of 16 clusters.  Therefore the value of the Hubble parameter derived in this study is not significantly affected by the adopted K-TFR slope.

It is important to note that the steeper slope produces mean distances to individual cluster members that are rotational velocity dependent. For example in the A400 cluster the 6 fastest and 6 slowest rotators have mean distance moduli of 34.84 +/-0.16 and 34.55 +/-0.27 respectively with a slope of -10.02. With the slope of -8.22 the mean distance moduli of the fastest and slowest rotators are 34.65 +/-0.15 and 34.55 +/-0.21 respectively suggesting that the shallower slope used in this study is more appropriate for a sample restricted to faster rotators. The discrepancy between fast and slow rotators is smaller than A400 for the Pisces supercluster (+0.12 mag) but even larger than A400 for the Coma cluster (+0.45 mag).

Table 8:  Cluster Hubble Parameter with K-TFR slope of -10.02

| Cluster | m-M K-TFR | $H_0$ |
|---------|-----------|-------|
| Pisces | 33.81 | 82.9 |
| A1367 | 34.47 | 83.8 |
| A400 | 34.69 | 83.4 |
| Coma | 34.77 | 84.0 |
| Cancer | 33.61 | 85.8 |
| Hydra | 33.35 | 87.2 |
| A2197/99 | 35.33 | 81.4 |
| A2634 | 35.08 | 83.7 |
| A3574 | 33.80 | 85.1 |
| A779 | 34.82 | 81.8 |
| A548 | 35.65 | 89.4 |
| A114 | 36.53 | 88.1 |
| A1736 | 35.47 | 93.3 |
| A3716 | 36.05 | 88.7 |
| ESO 596 | 34.82 | 76.5 |
| ESO 471 | 35.59 | 64.2 |

*4.3  Effect of inclinations*

In order to test the possible effects of galaxy inclinations on the derived value of $H_0$, the 218 ScI galaxies were grouped into 4 inclination bins (Table 9). It can be seen in Table 9 that the inclination bins from 50-69 degrees give a mean Hubble parameter of 84.3 km s$^{-1}$ Mpc$^{-1}$ – consistent with the weighted mean value found for the 16 galaxy clusters. The galaxies with inclinations from 50-69 degrees are especially important because they comprise two-thirds of the ScI sample and are least susceptible to the problems associated with low or high inclinations. The galaxies with inclinations from 35-49 degrees give $H_0$=79.7 km s$^{-1}$ Mpc$^{-1}$ which is mildly discrepant when compared with the other inclination bins. This discrepancy is not unexpected as uncertainty in corrected rotational velocities increases for galaxies closer to face on orientation.

Table 9: Hubble Parameter for 218 ScI galaxies group by inclination bins

| Inclination range (°) | N | $H_0$ |
|---|---|---|
| 35-49 | 38 | 79.7 |
| 50-59 | 75 | 84.3 |
| 60-69 | 70 | 84.3 |
| 70-80 | 35 | 83.1 |
| Total | 218 | 83.4 |

### 4.4 Effect of Visual inspection of images

As described in section 2, images of candidate galaxies for this sample were visually inspected to verify inclination estimates from axial ratios and morphological luminosity classification. For the cluster sample, galaxies not assigned a luminosity class in HyperLeda were in some cases found to have ScI group morphology (Fig 2a) and therefore had distances calculated with the ScI group zero point rather than the Sb group zero point. Thus within the clusters the effect of visual inspection of images leads to a slight increase in clusters distances if galaxies were incorrectly judged to have ScI group morphology.

For the ScI sample 140 of the 218 galaxies were classified as ScI group galaxies in HyperLeda. The remaining 78 galaxies were classified as ScI group galaxies from visual inspection of images. A sample of these galaxies is provided in Figure 4 to illustrate the identification of ScI morphology. Not surprisingly 61 of these 78 additions were at distances beyond 80.0 Mpc. The 78 added galaxies seem to have the effect of slightly increasing the overall value of $H_0$. The value of $H_0$ for the 140 galaxies classified as ScI group in HyperLeda is 82.2 +/-8 km s$^{-1}$ Mpc$^{-1}$ whereas the 78 galaxies added from visual inspection give $H_0$=85.7 +/-8 km s$^{-1}$ Mpc$^{-1}$.

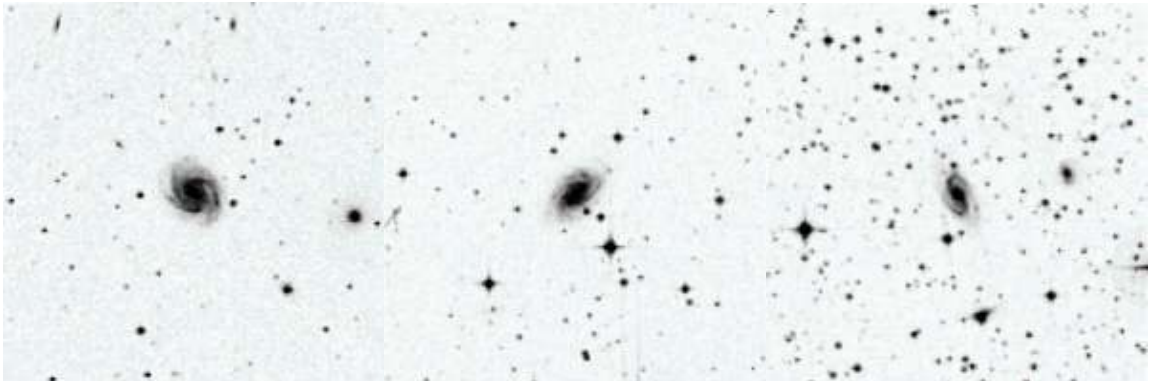

Figure 4a   ESO 578-16          Figure 4b   ESO 385-26          Figure 4c   ESO 328-16

However both subsets are very close to the weighted mean $H_0$ value found from the 16 clusters ($H_0$=84.2 km s$^{-1}$ Mpc$^{-1}$) and therefore support the higher value of $H_0$ found in this study.   It is also important to recognize that the galaxies added to the ScI sample from visual inspection of images would have even closer distances if the Sb zero point was utilized and therefore would give an even larger value for $H_0$.

*4.5  Effect of distance distribution rejection for clusters*

Table 10 lists the distances to the 16 clusters when the galaxies rejected as having anomalous K-TFR distances relative to the cluster mean are included.   The unweighted mean $H_0$ value then is 84.5 +/-5 km s$^{-1}$ Mpc$^{-1}$ which may be compared with an unweighted mean $H_0$ value of 84.9 +/-5 km s$^{-1}$ Mpc$^{-1}$ found when 25 galaxies are rejected from the clusters due to large discrepancies in their distances relative to the accepted galaxies.   The rejection of individual galaxies within the clusters based upon the distance distribution of the cluster members therefore has a negligible effect on the value of $H_0$.

*4.6  The Large Magellanic Cloud Distance Modulus*

In section 4.1 a Hubble parameter of 84 km s$^{-1}$ Mpc$^{-1}$ was found.   The zero point calibration for the K-TFR assumes a distance modulus to the LMC of 18.50 +/-0.10 as adopted by the HKP (Freedman et al 2001).   However, recent studies suggest the LMC distance modulus is closer to 18.39 +/-0.05 (Table 11).   Macri et al (2006) demonstrated an important metallicity effect in the Cepheid P-L relation and concluded that LMC distance modulus is 18.41 +/- 0.10.    A number of subsequent studies have adopted metallicity corrections for the Cepheid P-L relation.   Benedict et al (2007) provided new trigonometric parallaxes to Galactic Cepheids and found a distance modulus of 18.50 +/- 0.03 without metallicity corrections but 18.40+/-0.05 when applying the metallicity correction of Macri et al (2006).    van Leeuwen et al (2007) used revised Hipparcos parallaxes and found the distance modulus to the LMC is 18.52 +/-0.03 without metallicity corrections or 18.39 +/-0.05 with metallicity corrections.  An et al (2007) constructed the Cepheid P-L relation from 7 Galactic clusters with Cepheids and found a distance to the LMC of 18.48 without metallicity corrections and 18.34 +/-0.06 with metallicity corrections.  Fouque et al (2007) did not study the effects of metallicity on the P-L relation but concluded that the LMC distance modulus must be smaller than 18.50 after correcting for metallicity effects.

Table 10:   Clusters with distance distribution rejected galaxies included

| Cluster | N | m-M | Mpc | Vcmb | $H_0$ |
|---------|---|------|-------|-------|------|
| Template | | | | | |
| Pisces | 30 | 33.86 | 59.1 | 4716 | 79.8 |
| A1367 | 15 | 34.55 | 81.3 | 6618 | 81.4 |
| A400 | 16 | 34.62 | 83.9 | 7291 | 87.3 |
| Coma | 15 | 34.51 | 79.8 | 7402 | 92.8 |
| Cancer | 12 | 33.79 | 57.3 | 4534 | 79.1 |
| Hydra | 12 | 33.54 | 51.1 | 4144 | 81.1 |
| A2197/99 | 9 | 35.26 | 112.7 | 9479 | 84.1 |
| A2634 | 8 | 35.00 | 100.0 | 8689 | 86.9 |
| A3574 | 7 | 33.80 | 57.6 | 4903 | 85.1 |
| A779 | 5 | 34.76 | 89.5 | 7532 | 84.2 |
| Non-Template | | | | | |
| A548 | 8 | 35.64 | 134.3 | 12062 | 89.8 |
| A114 | 7 | 36.52 | 201.4 | 17810 | 88.4 |
| A1736 | 7 | 35.57 | 130.0 | 11585 | 89.1 |
| A3716 | 7 | 35.94 | 154.2 | 14382 | 93.3 |
| ESO 596 | 6 | 34.85 | 93.3 | 7043 | 75.5 |
| ESO 471 | 5 | 35.40 | 120.2 | 8429 | 70.1 |

Using Type II Cepheid's and RR Lyrae variables Feast et al (2008) find a LMC distance modulus of 18.37 +/-0.09. Catelan and Cortes provide a revised trigonometric parallax to RR Lyrae and derive a LMC distance modulus of 18.44 +/-0.11. From the K-band luminosity of the red clump in 17 LMC clusters Grocholski et al (2007) find the LMC distance modulus to be 18.40 +/-0.04.

It is clearly seen from these results that the latest studies indicate the best value for LMC distance modulus is ~0.10 mag smaller than the value adopted by the HKP when the required metallicity corrections are applied to the Cepheid P-L relation. The recent studies listed in Table 11 give an unweighted mean for the LMC distance modulus of 18.39 +/- 0.05 and require a downward revision of -0.11 mag for the zero points of the K-TFR calibrators in Tables 1 and 2. Applying this correction to the K-TFR distances of this study increases the derived value of the Hubble Parameter from the Type Dependent K-TFR to 88 +/-6 km s$^{-1}$ Mpc$^{-1}$.

Table 11: LMC distance modulus

| Study | m-M LMC | +/- |
|---|---|---|
| Macri et al 2006 | 18.41 | 0.10 |
| Benedict et al 2007 | 18.40 | 0.05 |
| van Leeuwen et al 2007 | 18.39 | 0.05 |
| Grocholski et al 2007 | 18.40 | 0.04 |
| An et al 2007 | 18.34 | 0.06 |
| Catelan & Cortes 2008 | 18.44 | 0.11 |
| Feast et al 2008 | 18.37 | 0.09 |

## 5. Comparison with the results of the Hubble Key Project

Since the value of the Hubble Parameter found in this analysis is significantly higher than the value found by Freedman et al (2001) from the I-TFR, SBF, FP, Type Ia SN, and Type II SN methods, the HKP results are reconsidered in the following sections. It is important to note that the HKP reported a Hubble parameter of 82 +/-9 from 11 clusters with Fundamental Plane distances – consistent with the result found in this study from 16 clusters and 218 ScI galaxies using the morphologically type dependent K-TFR. The remaining 4 methods used by the HKP give $H_0$=~71.

## 5.1   The HKP I-TFR

The HKP I-band TFR was calibrated by Sakai et al (2000). Final adjustments to cluster distances were presented in Freedman et al (2001) in order to account for the final Cepheid calibrator distances.   From the I-TFR the HKP found $H_0$=71 +/-7 km s$^{-1}$ Mpc$^{-1}$ – significantly smaller than the value found in this study.     Table 12 compares the distances to the 8 clusters in common with this study and Freedman et al (2001).     In every cluster except Coma, the HKP cluster distances are larger than the K-TFR distances found in this study with a mean distance modulus difference of +0.28 mag.

Figure 5 is a plot of the absolute magnitude ($M_K$) versus the logarithm of the rotational velocity (log Vrot) for the ScI group galaxies in the cluster sample (open circles) of this study at the I-TFR cluster distances determined by the HKP.  Also plotted in Figure 5 are the 36 ScI group calibrators from this study (filled circles).   It is clearly seen in Figure 5 that the cluster galaxies would be systematically over-luminous relative to the calibrators if they were at the I-TFR distances found by the HKP.

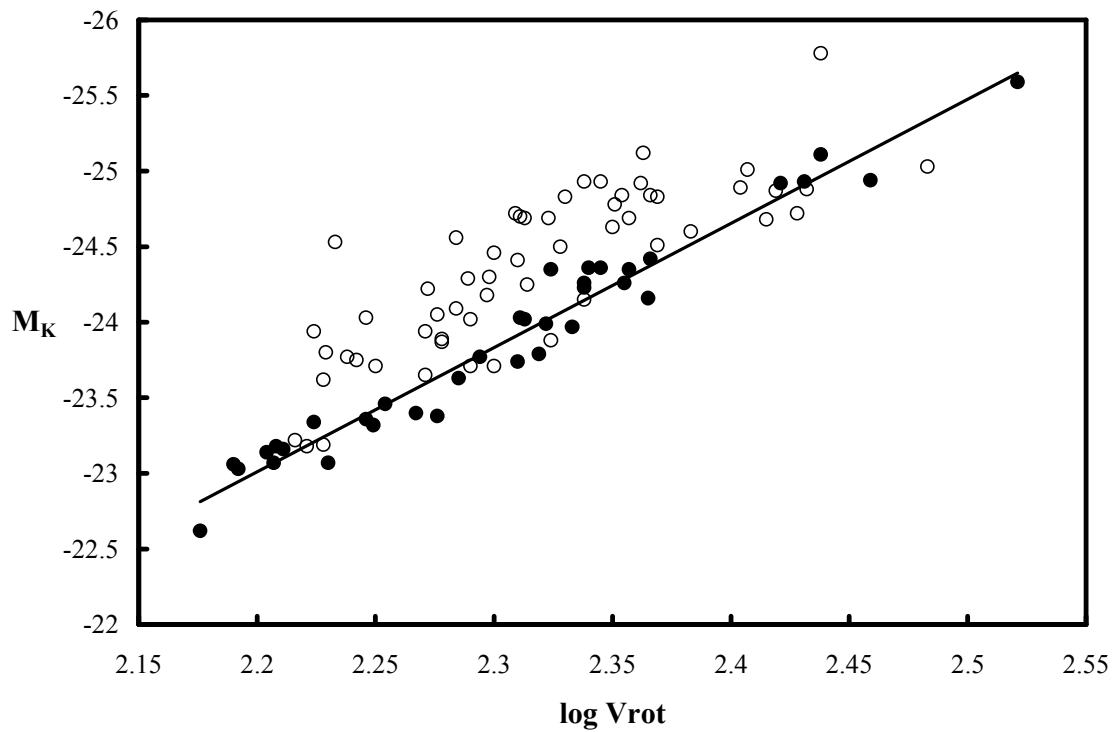

*Figure 5 – K-TFR plot for ScI galaxies in the cluster sample at the I-TFR distances reported by Freedman et al (2001).   Filled circles are the 36 calibrators from Tables 1 and 2 and solid line is a slope of -8.22.   Open circles are the ScI's in the cluster sample at the I-TFR distances of the HKP.*

There is independent evidence that the problem is not the K-TFR distances to the clusters found in this study, but rather the I-TFR distances found by the HKP. For example, the HKP found an I-TFR distance to Cen30 of 43.2 Mpc. However, a Cepheid distance to the Cen30 galaxy NGC 4603 was determined by Newman et al (1999) and the galaxy was found to be at 33.3 Mpc – 10 Mpc closer than the HKP I-TFR. Tonry et al (2001) found a distance to the Centaurus cluster of 32.8 Mpc from 9 early type galaxies with SBF distances - a value in excellent agreement with the NGC 4603 Cepheid distance.

The HKP I-TFR distance to the Antlia cluster is 45.1 Mpc. Four galaxies in the Antlia cluster with SBF distances in Tonry et al give a distance of 33.1 Mpc. Recently Bassino et al (2008) determined the distances to the giant ellipticals NGC 3258 and 3268 in Antlia from the globular cluster luminosity function(GCLF). The mean distance of the two giant ellipticals is 33.4 Mpc – in excellent agreement with the SBF distances. It should also be noted that 7 galaxies from the Antlia cluster were included in the calibration sample for the K-TFR of ScI galaxies using the Tonry et al SBF distances to Antlia (Table 2). The mean zero point of these 7 galaxies is in exact agreement (23.01 +/-0.16) with the overall mean found from the full sample of 36 ScI calibrators. In order for the HKP I-TFR distance to Antlia to be correct, the K-TFR zero point would need to be increased by 0.67 mag – which is $4.8\sigma$ larger than the observed scatter in the K-TFR zero point.

The explanation for the discrepancy in the HKP project I-TFR distances is beyond the scope of this paper. However, the above evidence strongly argues for a problem with the HKP I-TFR distances rather than the K-TFR distances found in this study.

Table 12: Comparison with HKP I-TFR distances

| Cluster | m-M K-TFR | m-M HKP I-TFR | I-TFR – K-TFR |
|---------|-----------|---------------|---------------|
| Pisces | 33.77 | 34.01 | +0.24 |
| A1367 | 34.49 | 34.75 | +0.26 |
| A400 | 34.60 | 34.73 | +0.13 |
| Coma | 34.70 | 34.66 | -0.04 |
| Cancer | 33.61 | 34.35 | +0.74 |
| Hydra | 33.43 | 33.83 | +0.40 |
| A3574 | 33.80 | 33.97 | +0.17 |
| A2634 | 35.00 | 35.30 | +0.30 |

*5.2  The HKP SBF distances*

The HKP found $H_0$=70 +/-6 km s$^{-1}$ Mpc$^{-1}$ from 6 galaxies with Surface Brightness Fluctuation distances.  For each of these galaxies it is possible to compare the HKP SBF distance with cluster or group FP, SBF, or K-TFR distance estimates from this or other studies.

*NGC 4881* is a member of the Coma cluster and for this galaxy the HKP finds a SBF distance of 102.3 Mpc.   The HKP FP distance to Coma is 85.8 Mpc and the K-TFR distance is 87.1 Mpc.   The SBF distance modulus to NGC 4881 is therefore +0.35 mag larger than found from other methods.   Considering that the FP distance is derived from 81 galaxies and the K-TFR distance is derived from 11 galaxies, it would be reasonable to conclude that NGC 4881 is simply on the backside of the Coma cluster and not representative of the mean Coma cluster distance.    In addition, Ferrarrese et al (2000) note that there is a large uncertainty in the NGC 4881 SBF distance because the galaxy was not observed in the V band.

*NGC 4373* (ESO 322-6) is a member of the Cen30 cluster and the HKP found a SBF distance of 36.3 Mpc for this galaxy.   This distance is in exact agreement with NGC 4373's nearby companion ESO 322-8 for which Tonry et al (2001) find a SBF distance of 36.3 Mpc.

*NGC 708* is a member of the Abell 262 cluster and the HKP reports a SBF distance of 68.2 Mpc.   The K-TFR distance to A262 is 56.5 Mpc from 10 galaxies in A262 meeting the ScI or Sb/ScIII group selection criteria of this study.   The NGC 708 SBF distance modulus is 0.41 mag larger than the K-TFR distance modulus.

*NGC 5193* is listed as a member of the Abell 3560 cluster in Ferrarese et al (2000). However, Abell 3560 is actually a background cluster with cz= ~15000 km s$^{-1}$.   NGC 5193 is coincident with coordinate distribution of the Abell 3574 cluster sample of Springob et al (2007).   The HKP reports a SBF distance of 51.5 Mpc for NGC 5193 in excellent agreement with the HKP FP distance to A3574 of 51.6 Mpc.   The K-TFR distance to Abell 3574 is 57.6 Mpc - somewhat larger than the SBF and FP distance. Corrected to the Cosmic Microwave Background reference frame the redshift of NGC 5193 is 3991 km s$^{-1}$ which is consistent with the lower redshift members of Abell 3574.

IC4296 is also coincident with the Abell 3574 cluster and has a SBF distance of 55.5 Mpc – very close to the value found with the K-TFR and consistent with the HKP FP distance.

*NGC 7014* has a SBF distance of 67.3 Mpc.    The ScI galaxy ESO 286-79 is the closest neighbor to NGC 7014 for which a K-TFR distance can be calculated.   Using the rotational velocity provided by MF96 (log Vrot = 2.468) and a corrected 2MASS K-band magnitude of 8.71, the K-TFR distance to ESO 286-79 is 60.8 Mpc.   The NGC 7014 SBF distance modulus is larger by +0.22 mag.

The HKP SBF estimate for $H_0$ is based upon only 6 galaxies in 6 clusters.  There is a certain amount of risk in determining the distance to a cluster from a single galaxy as the selected galaxy may be on the front or backside of the cluster.  It is therefore reasonable to conclude that the value of $H_0$ found in this study with the K-TFR using a sample of 16 galaxy clusters with a minimum of 5 cluster members and a sample of 218 ScI galaxies provides a more reliable sampling of the Hubble flow than a value derived from 6 SBF distances.

However, it is interesting to note that the three nearest galaxies in the HKP SBF sample (NGC 4373, IC 4296 and NGC 5193) have distances in excellent agreement with Tonry et al (2001) SBF, HKP FP, and K-TFR distances from this study.   The three most distant SBF galaxies have distance moduli a mean +0.33 mag larger than the distance moduli found from the other methods.  This may indicate a systematic problem with the application of the SBF method at distances beyond ~ 60 Mpc, but a larger number of SBF distances will be needed in order to test that possibility in any meaningful way.

*5.3  Type II and Type Ia SN*

The HKP found $H_0$=72 +/-9 km s$^{-1}$ Mpc$^{-1}$from 4 Type II SN.   Four galaxies is too small a sample to draw meaningful conclusions about the value of $H_0$.   In addition, only 3 galaxies with Cepheid distances were available for fixing the zero point of the Type II SN distance scale (Table 11 of Freedman et al 2001).    Given these problems with sample size, the HKP Type II SN result is not considered here.

The HKP found $H_0$=71 +/-6 km s$^{-1}$ Mpc$^{-1}$ from 36 Type Ia SN. In compiling the K-TFR calibrator sample for this study 7 ScI galaxies were identified as nearby companions to 4 of the Type Ia SN. For the remaining 32 Type Ia SN, companion ScI galaxies for which K-TFR distances could be calculated were not found. The 7 ScI's with calibration distances defined by the Type Ia SN are shown as open diamonds against the remaining 29 calibrators (filled circles) in Figure 6. Note that the 7 galaxies with Type Ia SN distances fall tightly on the mean relation defined by the other 29 galaxies (solid line). It should also be recalled from section 2.1 that the slope defined by the 7 ScI's with SN calibration distances is -8.16 which is in excellent agreement with the slope defined by the full sample of 36 calibrators. In addition the mean zero point of these 7 galaxies is 23.01 +/-0.06 – in exact agreement with the overall mean.

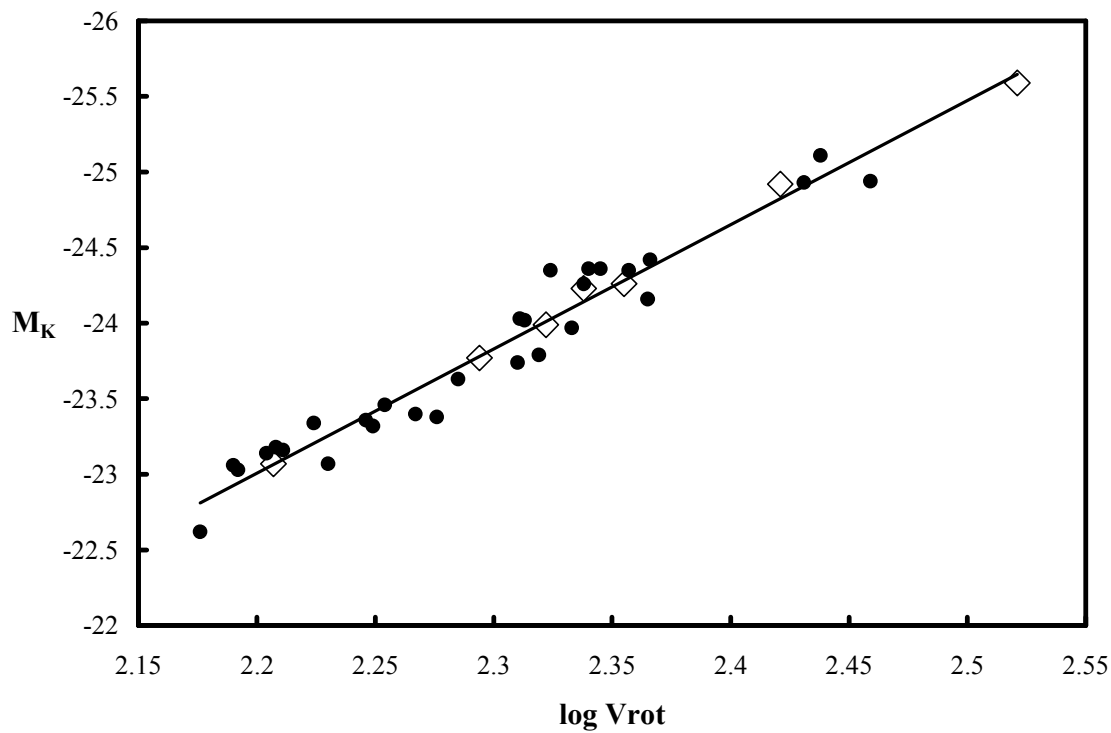

*Figure 6 – Comparison of 7 ScI calibrators (open diamonds) that are companions of galaxies with Type Ia SN distances compared with the remaining 29 ScI calibrators (filled circles). Solid line has a slope of -8.22.*

While the comparison sample is an uncomfortably small sample of only 4 Type Ia SN distances, the comparison suggests that there is excellent agreement between the Type Ia SN distances and the K-TFR distances of this study. The comparison can only be improved when greater numbers of Type Ia SN distances become available in the local universe.

*5.4    Can the Hubble Parameter be 70 km s$^{-1}$ Mpc$^{-1}$?*

Supported by the results of the HKP (Freedman et al 2001) and WMAP (Spergel et al 2003,2006; Hinshaw et al 2009; Dunkley et al 2009) it is generally accepted that the value of $H_0$ is ~70 km s$^{-1}$ Mpc$^{-1}$. Sandage et al (2006) find an even smaller value for $H_0$ of 62+/-5 km s$^{-1}$ Mpc$^{-1}$. In this study, the value of $H_0$ was found to be $H_0$ = 84 +/-6 km s$^{-1}$ Mpc$^{-1}$ when adopting the HKP zero point for the LMC. Adopting a LMC distance modulus of 18.39 indicated from recent studies (section 4.6) increases $H_0$ to 88+/-6 km s$^{-1}$ Mpc$^{-1}$. The values found in this study are significantly larger than those found in other studies.

Figure 7 is a plot of the K-band absolute magnitudes vs. the logarithm of the rotational velocity for the 36 calibrators in Tables 1 and 2 and the ScI galaxies in the 16 clusters. The K-band absolute magnitudes for the calibrators are calculated for a LMC distance modulus of 18.39 whereas absolute magnitudes of the cluster galaxies are derived assuming the mean cluster distances for $H_0$=70 km s$^{-1}$ Mpc$^{-1}$. A Hubble Parameter of 70 km s$^{-1}$ Mpc$^{-1}$ is adopted as a reasonable average of the $H_0$ values the HKP and Sandage teams would find using a LMC distance modulus of 18.39.

It is evident in Figure 7 that the K-TFR defined by the calibrators is inconsistent with $H_0$=70 km s$^{-1}$ Mpc$^{-1}$. There are reasons for suspecting the discrepancy between the HKP results and this study could arise from problems with the HKP distances and samples. First, it is important to note again that the HKP value of $H_0$ from the Fundamental Plane was $H_0$=82 +/-9 km s$^{-1}$ Mpc$^{-1}$ which is in excellent agreement with the results of this study.

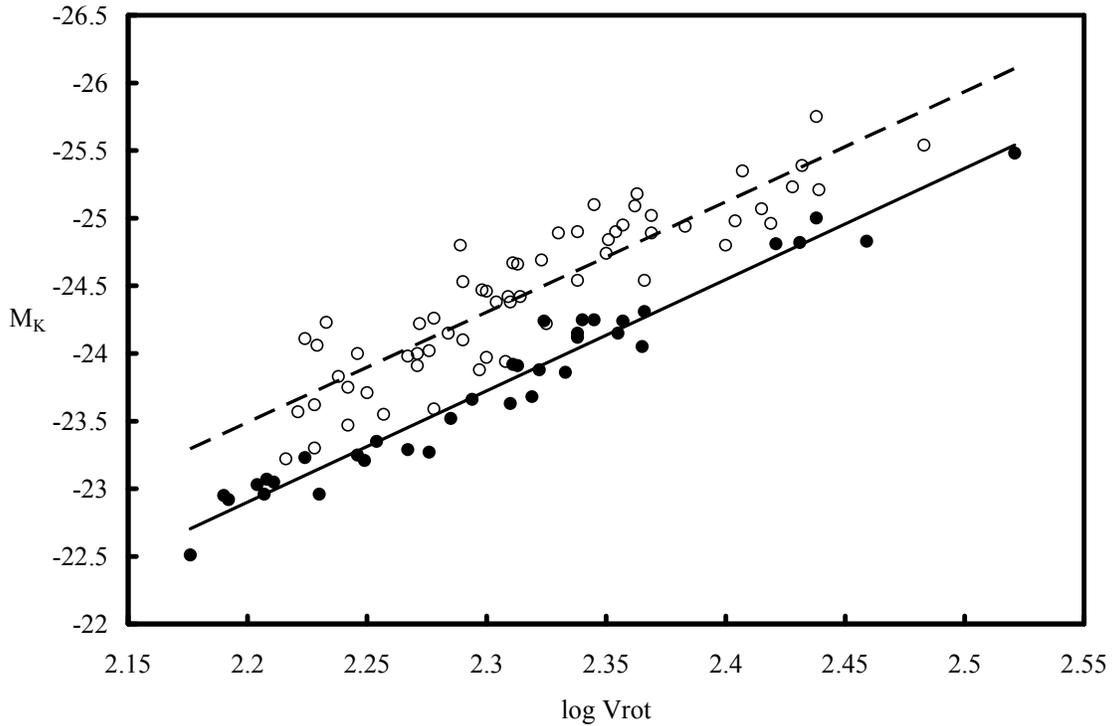

*Figure 7 – K-TFR plot for ScI galaxies in the 16 clusters (Table 6) at redshift distances using $H_0=70$ km s$^{-1}$ Mpc$^{-1}$. Open circles are cluster ScI's at Hubble distances and dashed line is a least squares fit. Filled circles are the 36 calibrators from Tables 1 and 2 with a LMC distance modulus of 18.39 and solid line is a least squares fit. The ScI's are systematically more luminous than the calibrators at the $H_0=70$ distances.*

It was shown in section 5.1 that the HKP I-TFR distances are significantly at odds with SBF, Cepheid, and GCLF distance estimates for the Antlia and Cen 30 clusters whereas the K-TFR calibration of this study is an excellent fit to the other distance estimates. In addition the K-TFR selection criteria used in this sample eliminates galaxies likely to contribute to Malmquist bias and large distance errors – whereas a more generous sample selection criteria is employed by Sakai et al (2000).

The HKP SBF $H_0$ value was determined from only 6 individual SBF distance estimates and the HKP Type II SN $H_0$ estimate was determined from only 4 distance estimates with 3 zero point calibrators. These sample sizes used to determine $H_0$ are 6-9 times smaller than the zero point calibrator sample used in this study and do not compare favorably with the K-TFR sample sizes used to calculate $H_0$ in this study. In addition the SBF distances agree well with FP and K-TFR distance estimates for the closest galaxies, but are systematically too large for the more distant galaxies in the HKP sample suggesting a possible problem with the SBF at larger distances.

Finally, the HKP used 36 Type Ia SN for which a small comparison sample suggests excellent agreement between the SN distances and the K-TFR calibration of this study. However, the HKP again finds a much smaller value of $H_0$ than found in this study. The reason for this discrepancy will only be resolved with a larger sample of galaxies with both Type Ia SN and K-TFR distances. However, the discrepancy suggests that either the HKP Type Ia SN sample or the K-TFR sample of this study inadequately samples the Hubble flow. Given that the HKP Type Ia SN sample is significantly smaller – and spread over a much larger distance range (58.0 to 467.0 Mpc) than the ScI and cluster samples of this study (40.0 to 140.0 Mpc range for all but 2 clusters), it is possible that the Type Ia SN distances have not uniformly sampled the Hubble flow.

## 6. Implications for Cosmology

The value of the Hubble Constant provides an important constraint upon cosmological models. The HKP value for $H_0$ has been argued to support the $\Lambda$-CDM concordance cosmology model (hereafter $\Lambda$-CDM – e.g. Spergel et al 2003, 2006; Hinshaw et al 2009; Dunkley et al 2009; Komatsu et al 2009). In this section we briefly consider implications for cosmology if $H_0 = 84$ km s$^{-1}$ Mpc$^{-1}$.

A basic requirement of any cosmological model is that the Universe must be older than the oldest objects contained within it. Currently, the oldest dated objects in the Milky Way are globular clusters which have ages as large as $\sim 13$ Gyr (e.g. Salaris & Weiss 2002, Rakos & Schombert 2005). If it is assumed that the Milky Way is nearly as old as the Universe, then any internally consistent set of cosmological parameters must be able to account for a Universe that is at least 13.5 Gyr.

Retaining the prevailing $\Lambda$-CDM parameters with a flat universe, $\Omega m = 0.27$ and $\Omega \Lambda = 0.73$ a Hubble constant of 84 km s$^{-1}$ Mpc$^{-1}$ results in a universe that is 11.55 Gyr (Wright 2006) - younger than the oldest globular clusters. This discrepancy can be resolved at the cost of a lower matter density and higher dark energy density. A flat universe with $H_0$=84, $\Omega m = 0.14$ and $\Omega \Lambda = 0.86$ would have an age of 13.71 Gyr (Wright 2006) and would be consistent with the ages of the oldest globular clusters. However, a matter density of, $\Omega m = 0.14$ is only at best marginally consistent with the matter density estimated from galaxy cluster studies. Carlberg et al (1996) found $\Omega m = 0.24$ +/-0.05 and more recently Muzzin et al (2007) find $\Omega m = 0.22$ +/- 0.02.

This discrepancy is made more serious if other galaxies are older than the Milky Way thus requiring an even older age for the universe. For example, Lee et al (2001) found that the globular clusters in NGC 1399 may be several billion years older than the Galactic GC system. Bregman, Temi, & Bregman (2006) determined the ages of 29 elliptical galaxies with IR spectral energy distributions and found 8 galaxies (27.6% of their sample) had ages greater than 15.7 Gyr with the oldest galaxy being 20.6 Gyr. Based upon the obvious disagreement with $\Lambda$-CDM parameters, Bregman et al argued that this discrepancy might indicate a problem for the absolute accuracy of their age dates and shifted the ages in their sample into a range that accommodates the standard $\Lambda$-CDM model. It is noted here that with $H_0$=84, the Bregman et al ages would require a matter density of $\Omega m = 0.06$ to accommodate elliptical galaxies with ages of 16 Gyr. This matter density is clearly inconsistent with that found from galaxy clusters (Carlberg et al 1996; Muzzin et al 2007).

It is of concern to note that it is not possible to simultaneously reconcile the observed matter density of the universe ($\Omega m = 0.22$ +/- 0.02), estimated ages of the oldest globular clusters (~13 Gyr), and value of the Hubble Parameter found in this study ($H_0$=84+/-6 km $s^{-1}$ $Mpc^{-1}$) with $\Lambda$-CDM cosmology. Note that the discrepancy with $\Lambda$-CDM expectations becomes even more severe when the latest results for the LMC distance modulus are taken into account because the value of $H_0$ found in this study is then raised to 88 +/-6 km $s^{-1}$ $Mpc^{-1}$. Whether or not this indicates a problem for the standard cosmological model will require further investigation.

## 7. Conclusion

The morphologically type dependent Tully-Fisher Relation (Russell 2004) was calibrated in the Ks-band for galaxies with a minimum rotational velocity of 150 km s$^{-1}$. Distances were derived for galaxies in 16 galaxy clusters and 218 ScI galaxies using rotational velocities from the Springob et al (2007) database. Applying unweighted and weighted means as well as binning of sample galaxies by distance and inclination, the value of the Hubble Parameter was consistently found to fall in the range of 82 to 85 km s$^{-1}$ Mpc$^{-1}$ with a preferred value of 84 +/-6 km s$^{-1}$ Mpc$^{-1}$. If recent results for the value of the LMC distance modulus are adopted, the value of $H_0$ would increase to 88 +/-6 km s$^{-1}$ Mpc$^{-1}$.

It is very difficult to fit the observed matter density of the universe derived from galaxy clusters ($\Omega m = 0.22$ +/- 0.02 – Muzzin et al 2007); ages of the oldest globular clusters (~13 Gyr – Salaris&Weiss 2002); and the value of $H_0$ found in this study ($H_0$=84 +/- 6) with standard $\Lambda$-CDM cosmology. In order to reconcile the age of the universe for a flat universe with dominant dark energy component and $H_0$=84 requires a matter density of $\Omega m = 0.14$ – which is 4.8$\sigma$ below the value found by Muzzin et al (2007).

While a Hubble Parameter of 84 km s$^{-1}$ Mpc$^{-1}$ is significantly different from the value found by Freedman et al (2001), it was shown that the Freedman et al I-TFR distances to clusters result in a K-TFR for ScI galaxies in clusters that is systematically overluminous by ~0.35 mag relative to the K-TFR defined by 36 calibrators in this study. Since the HKP I-TFR distances are also inconsistent with SBF, Cepheid, and GCLF distances to the Antlia and Cen30 clusters, it was concluded that the problem most likely lies with the HKP I-TFR distances rather than the K-TFR distances derived in this study. The discrepancy between the HKP I-TFR results and the K-TFR results of this study might seem puzzling in light of the fact that the 36 calibrators of this study are fixed to the same Cepheid distance scale utilized by the HKP. However, sample selection criteria play an important role, and evidence was presented that the strict selection criteria of this study provide more reliable distances – as well as a slope more appropriate for the faster rotators used in this study.

A Hubble Parameter of 84 km s$^{-1}$ Mpc$^{-1}$ is also inconsistent with the recent best estimate from WMAP (Hinshaw et al 2009; Dunkley et al 2009; Komatsu et al 2009) which finds $H_0$=71.9 +/-2.7 km s$^{-1}$ Mpc$^{-1}$ assuming a 6 parameter $\Lambda$-CDM cosmology with flat geometry.    The implications of this discrepancy for cosmological modeling will require further study and it is concluded that the value of $H_0$ found in this study does not confirm the WMAP value of $H_0$.

**Acknowledgements**

I would like to thank the many researchers that have collected the data utilized in this research.  This research has made use of the NASA/IPAC Extragalactic Database (NED) which is operated by the Jet Propulsion Laboratory, California Institute of Technology, under contract with the National Aeronautics and Space Administration.  This research has made use of the HyperLeda database (http://leda.univ-lyon1.fr).  Figures 2 and 4 are images extracted from the Digitized Sky Survey. The Digitized Sky Surveys were produced at the Space Telescope Science Institute under U.S. Government grant NAG W-2166. The images of these surveys are based on photographic data obtained using the Oschin Schmidt Telescope on Palomar Mountain and the UK Schmidt Telescope.This research also makes use of data from Two Micron All Sky Survey, which is a joint project of the University of Massachusetts and the Infrared Processing and Analysis Center, funded by NASA and the NSF.   I would like to thank an anonymous referee for suggestions that led to significant improvements in the presentation of these results.